\documentclass[a4paper,11pt]{article}
\pdfoutput=1 

\usepackage{jinstpub} 

\usepackage{lineno}

\usepackage{overpic}

\title{\boldmath Research on proton beam spot imaging based on pixelated gamma detector}

\author[a, b]{Y. K. Sun,}
\author[a, b, 1]{H. T. Jing \note{Corresponding author.}}
\author[a,b, c]{B. B. Tian,}
\author[a,b, d]{X. L. Gao,}
\author[a, b, e]{X. Y. Yang}

\affiliation[a]{Institute of High Energy Physics, Chinese Academy of Sciences (CAS), Beijing, 100049, China}
\affiliation[b]{Spallation Neutron Source Science Center (SNSSC), Dongguan, 523803, China}
\affiliation[c]{School of Energy and Power Engineering, Xi'an Jiaotong University, Xi'an, 710049, China}
\affiliation[d]{Hebei Normal University, Shijiazhuang, 10094, China}
\affiliation[e]{Guangxi Normal University, Guilin, 541004, China}

\emailAdd{jinght@ihep.ac.cn, sunyankun@ihep.ac.cn}

\abstract{The primary secondary particles from the spallation target of the China Spallation Neutron Source are mainly gammas and neutrons, which are related to the distribution of the incident proton. The reconstruction of proton beam spot could be implemented based on the distribution of secondary particles. The methods of pinhole imaging and Compton imaging are developed by measuring the gamma distribution based on the pixelated detector. The secondary gammas could be detected by the pixelated gamma detector directly. The neutron can be identified by detecting the characteristic (478 keV) $\gamma$-rays from the $^{10}$B($n, \alpha$)  reactions. In order to detect secondary neutrons, a layer of $^{10}$B converter is added before the pixelated gamma detector. The pixelated gamma detector is sensitive to the characteristic (478 keV) $\gamma$-rays and then the neutron imaging could be achieved based on measuring the distribution of the characteristic gamma. }


\begin{document}
\maketitle
\flushbottom

\setpagewiselinenumbers

\section{Introduction}

It is crucial to monitor beam status by measuring the parameters of the beam profile. Although the Faraday cup generally is used for measuring the beam intensity,  the pixelated one could measure the beam profile~\cite{PixelatedFaradayCup}. The wall current monitor system can provide the information of the longitudinal profile about the proton bunches~\cite{WallCurrentMonitorKEK, WallCurrentMonitorRHIC}. The two-dimensional (2-D) transverse profile of proton beam in the Linac can be measured with a multi-wire scanner (MWS) in a no-destructive way~\cite{MultWireScanner}. One kind of stripline-type beam position monitor (BPM) was designed and used for measuring the position and phase of the proton beam in air in a non-destructive~\cite{StriplineBPM}. A proton imaging system has been also developed to measure the proton range, which is composed of a scintillator and a charge-coupled device~\cite{CCDScintillator}.

\begin{figure}[htbp]
\begin{center}
\begin{overpic}[width=10.0cm,height=4.5cm,angle=0]{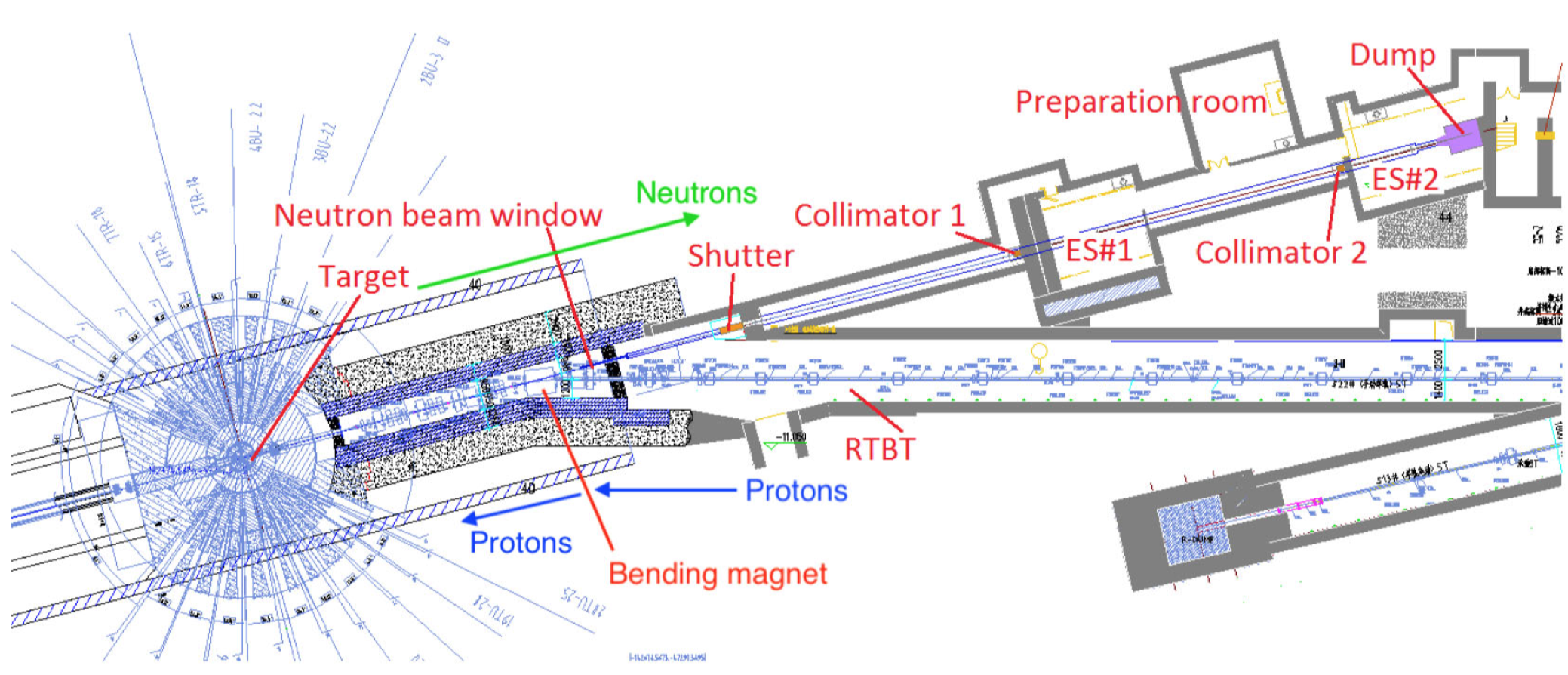}
\end{overpic}
\end{center}
\caption{Schematic diagram of the Back-n beam line at CSNS. }
\label{CSNSWNS}
\end{figure}

China Spallation Neutron Source (CSNS) has been built and commissioned successfully since August 2018, whose purpose is dedicated to the multidisciplinary research on material science~\cite{CSNS1}. It consists of an 80 MeV linear proton accelerator, a 1.6 GeV proton Rapid Cycling Synchrotron (RCS), a target station and the different kinds of neutron spectrometers. For example, an associated white neutron beam line was built for nuclear data measurement by exploiting the back-streaming neutrons (Back-n)~\cite{WNS1, WNS2}. Figure~\ref{CSNSWNS} shows the schematic diagram of the Back-n beam line. The proton distribution from the rapid cycling synchrotron at CSNS is measured by the position sensitive MWS~\cite{MultWireScannerCSNS, MultWireScannerCSNS2}. Because the primary secondary gammas and neutrons are related to the distribution of the incident proton, there is an opportunity to measure the proton beam spot of the spallation target on the Back-n beam line.

In this work, we used Geant4 software~\cite{Geant4} to simulate the physical process of generating neutrons and gammas by slamming protons onto a cylindrical tungsten target. The gamma distribution on the target surface was measured by the pixelated detector of Cadmium Zinc Telluride (CZT). The proton beam distribution is reconstructed by utilizing the gamma distribution based on the method of pinhole imaging. In addition, when the pixelated gamma detector could be combined with a neutron converter of $^{10}$B, the method of Compton imaging is used for neutron imaging based on measuring the characteristic (478 keV) $\gamma$-rays so that we can use neutron distribution to accomplish the reconstruction of proton beam spot.

\section{The principle of Compton imaging and pinhole imaging}

\begin{figure}[htbp]
\begin{center}
\begin{overpic}[width=4.6cm,height=4.4cm,angle=0]{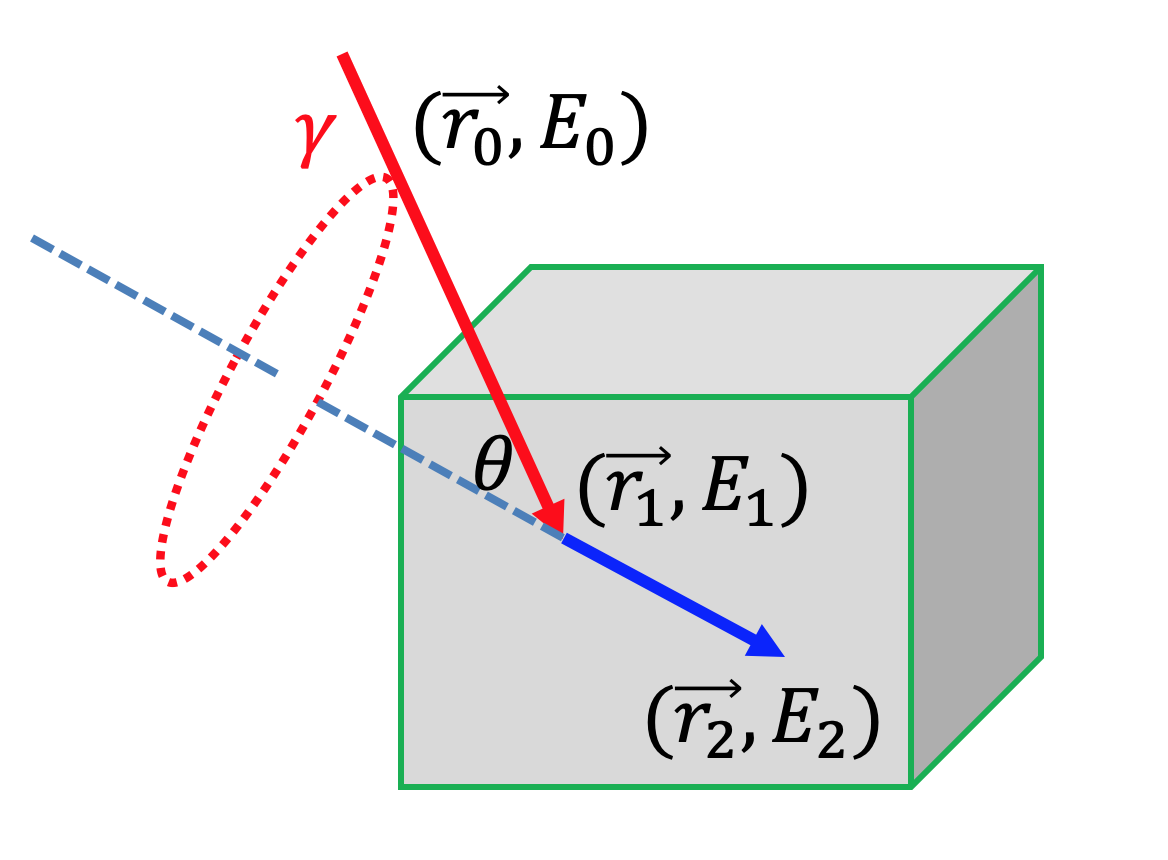}
\end{overpic}
\begin{overpic}[width=4.5cm,height=4.cm,angle=0]{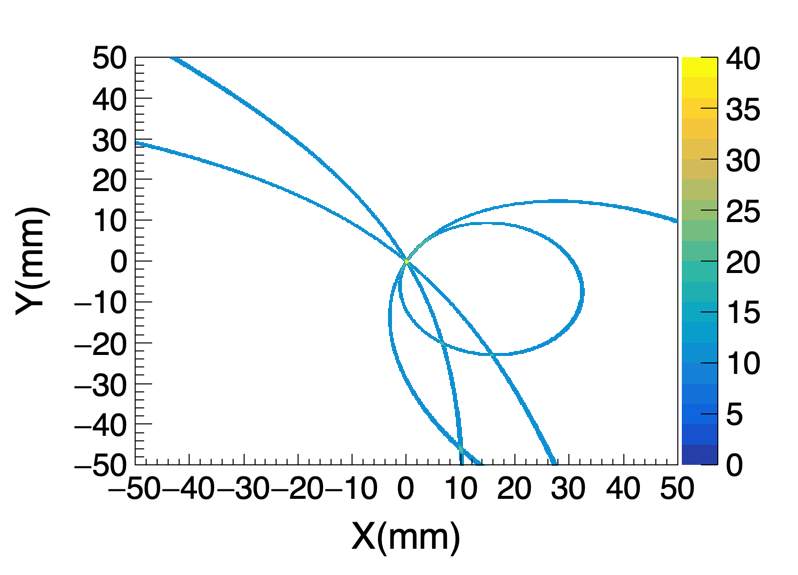}
\end{overpic}
\begin{overpic}[width=4.5cm,height=4.cm,angle=0]{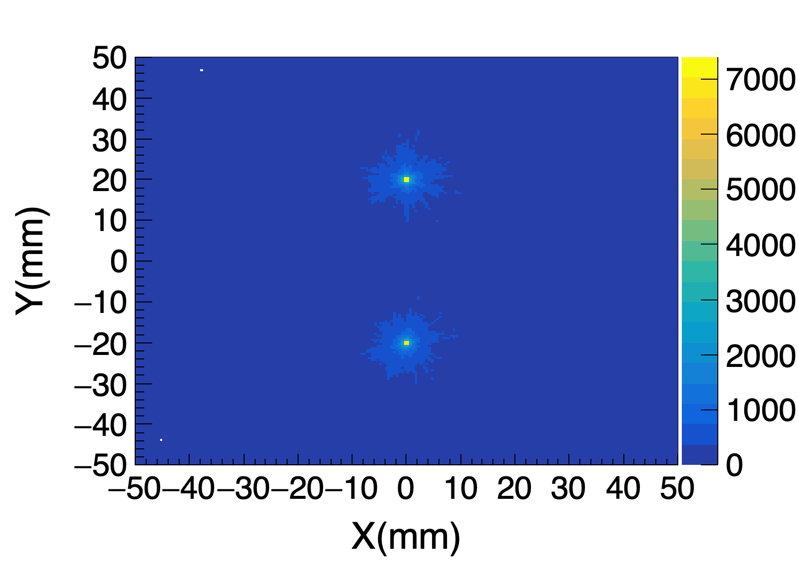}
\end{overpic}
\end{center}
\caption{(Left) Schematic diagram of Compton imaging for a single event. (Middle) The oval projection of the cones from four events of Compton scattering process. (Right) Compton imaging of two gamma sources, the distance between two gamma sources is 40 mm. The center of double gamma sources is 50 mm away from the pixelated CZT detector.}
\label{ComptonProcess}
\end{figure}

Compton imaging is a promising approach for the localization of $\gamma$-ray source. The process of Compton scattering is illustrated in Fig~\ref{ComptonProcess} (Left). The back-projection (BP) algorithm~\cite{BackProjection}  and maximum likelihood estimation (MLE) algorithm~\cite{MLM, LMLM} are used to Compton imaging. The reconstruction technique of Compton imaging in this work is based on the simple BP algorithm~\cite{ComptonImage1, 3DCZT}. The distance-weighted of Compton scattering angles is defined as

\begin{center}
\begin{equation}
Distance =  \frac{1.0}{[cos\theta(\vec {r}_{1}, \vec {r}_{2})  - cos\theta(E_{1}, E_{2}) ]^{2}}.
\label{DistanceWeighted}
\end{equation}
\end{center}

One can determine the scattering angles by using the information of the position and deposited energy of gamma respectively. The equations are described as

\begin{center}
\begin{equation}
cos\theta(\vec {r_{1}}, \vec {r_{2}}) =  \frac{\vec {r}_{1} \cdot \vec {r}_{2} }{|\vec {r}_{1} | \cdot |\vec {r}_{1} |},
\label{ComptonFuncPosition}
\end{equation}
\end{center}

\begin{center}
\begin{equation}
cos\theta(E_{1}, E_{2}) =  1 + \frac{m_{e}c^{2}}{E_{1} + E_{2}} - \frac{m_{e}c^{2}}{E_{2}}.
\label{ComptonFuncEnergy}
\end{equation}
\end{center}

Where $E_{1}$ is the deposited energy of the recoiling electron, $E_{2}$ is the energy of the scattering gamma, $m_{e}$ is the normal mass of the electron, ${c}$ is the speed of the light. $\vec {r}_{1}$ is the vector from $r_{0}$ to $r_{1}$ and $\vec {r}_{2}$ is the vector from $r_{1}$ to $r_{2}$, $r_{0}$ is the position of gamma source, $r_{1}$ and $r_{2}$ represent the three-dimensional (3-D) position of each radiation interaction in the detector.

A Monte Carlo (MC) sample of Compton process was generated by using GEANT4 software~\cite{Geant4}. The size of the pixelated CZT detector is 25.4 mm$\times$25.4 mm$\times$5 mm. The distance between two gamma sources is 40 mm. The single gamma source and the center of double gamma sources is 50 mm away from the CZT detector. Compton imaging was reconstructed with a simple BP method. The position of 478 keV point-liked gamma source could be located to the exterior surface of a cone by calculating the distance-weighted for each interaction event. The true information of positions and the energy obtained from the simulation is used for calculating the cosine values. Figure~\ref{ComptonProcess}(Middle) illustrates the oval projection of cones in a rectangular coordinate system from four events. The intersections of the oval projection indicate the position of two gamma source, as shown in Fig.~\ref{ComptonProcess}(Right).

Pinhole imaging relies on the principle of the rectilinear theory of light and appears upside down in the physical image space. It is a practical device of offering freedom from distortion and virtually infinite depth of field~\cite{PhiholeOptics}. A pinhole imaging diagnostic has been implemented to measure the electron beam position and profile in the SPEAR storage ring~\cite{PinholeImagingForBeam}. The imaging system of variable and moving pinhole arrays based on a time multiplexing method achieves much better resolution and signal-to-noise~\cite{PinholeImagingForOptical}. Because the proton beam has a high intensity, pinhole imaging is suitable for measuring the proton beam spot on the target. Compton imaging has an advantage of detecting efficiency so that it can be applied to the neutron imaging.

\section{The feasibility of realization for Compton imaging and pinhole imaging}

The gamma detector has to provide the information of the position and deposited energy of gamma according to the principle of Compton imaging. The position and energy resolution of the detector system affect the angular resolution of Compton imaging. The semiconductor detectors, such as high-purity germanium (HPGe), have higher detection efficiency and energy resolution compared to the inorganic crystals and plastic scintillators. The HPGe is the semiconductor detector with the best energy resolution. Due to the narrow band gap, the detector of HPGe only works in the liquid nitrogen temperature zone. The CZT has a wide band gap and can work at room temperature. Therefore, the Compton cameras have been developed based on the 3D position-sensitive semiconductor detector~\cite{H3D1, H3D2, H3D3, H3D4}. The positional resolution of the interaction points in the x and y directions was limited by the finite size of the CZT, and that in the z direction was limited by the finite resolution of drift time. The best angular resolution of the Compton camera consisting of a single Timepix3 detector with a thick 2 mm CdTe sensor reaches the order of a few degrees~\cite{BestComptonImage1, BestComptonImage2}. Figure 3 shows the promising applications of proton imaging on the target surface. Comparing with measuring the secondary gamma distribution, a converter of 10B is required for the secondary neutron detection.

\begin{figure}[htbp]
\begin{center}
\begin{overpic}[width=15.0cm,height=5.5cm,angle=0]{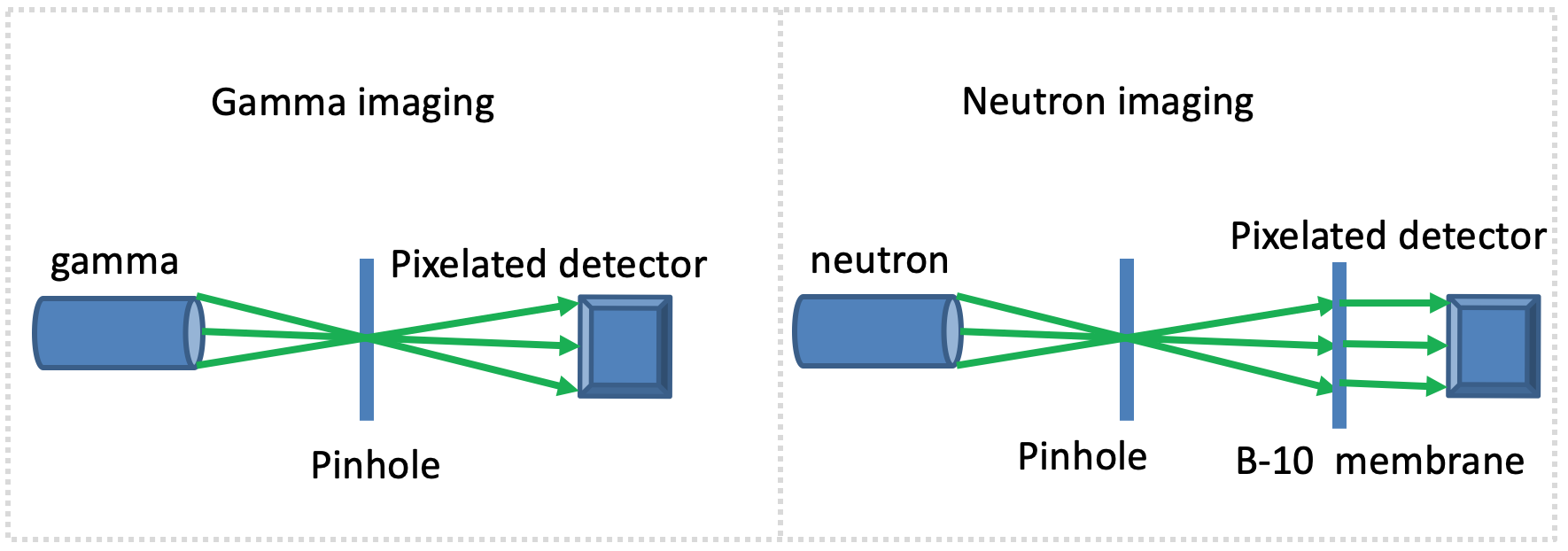}
\end{overpic}
\end{center}
\caption{Scheme of the experimental setup of proton beam spot based on the distribution of the secondary particles. (Left) Gamma imaging. (Right) Neutron imaging.}
\label{DesignOfExperiment}
\end{figure}

\subsection{Gamma imaging}

\begin{figure}[htbp]
\begin{center}
\begin{overpic}[width=7.5cm,height=5.5cm,angle=0]{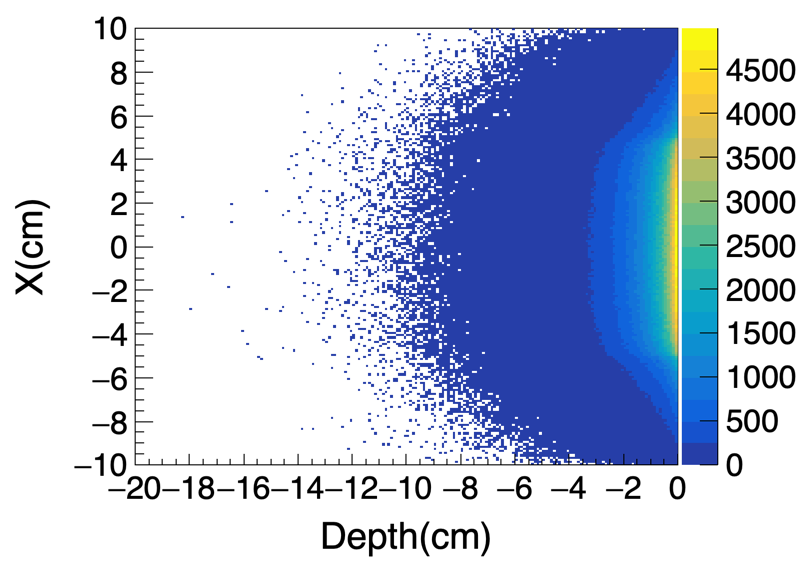}
\end{overpic}
\begin{overpic}[width=7.5cm,height=5.5cm,angle=0]{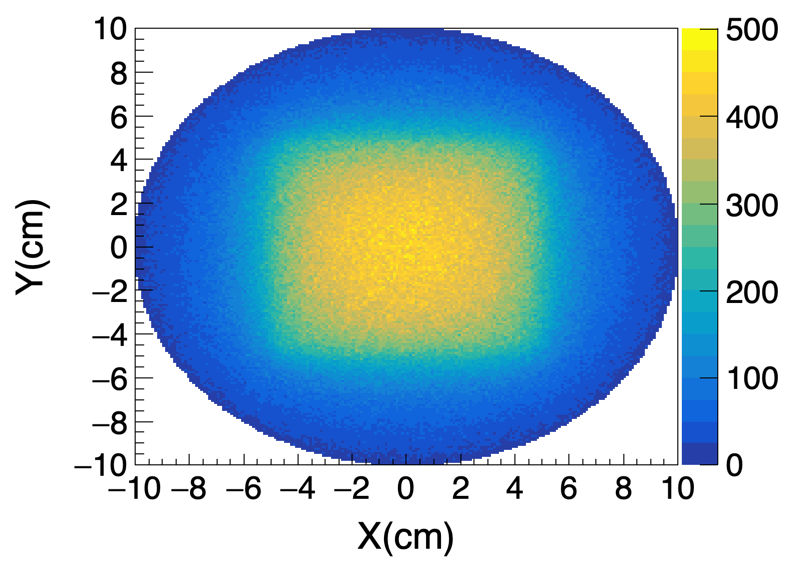}
\end{overpic}
\end{center}
\caption{(Left) Vertex position of secondary gammas. (Right) The 2-D distribution of gammas on the target surface.}
\label{SecondaryParticles}
\end{figure}

The physics list of QBBC from GEANT4 was used to simulate the interaction of 1.6 GeV proton beam on a cylindrical tungsten target with a radius of 10 cm and a height of 30 cm. The incident proton beam is uniformly square distribution with the length of a side is 5 cm. The secondary particles of neutron and gamma were emitted in the backward direction. The vertex position of secondary gammas could be found in Fig.~\ref{SecondaryParticles} (Left). Due to the relationship between the distribution of incident protons and the distribution of back-streaming secondary gamma, we utilize the distribution of secondary gamma to reconstruct the distribution of proton beam spot on the target surface. Figure ~\ref{SecondaryParticles} (Right) shows the 2-D position of secondary gamma on the target surface.

The distribution of proton beam spot is reconstructed according to the linear equation that is defined as

\begin{center}
\begin{equation}
\vec{G}(x, y) = \widehat{M} \times \vec{P}(x, y),
\label{LinearEquation}
\end{equation}
\label{LinearEquation}
\end{center}

where $\vec{G}(x, y)$ is 2-D gamma distribution, $\widehat{M}$ is the response matrix, and $\vec{P}(x, y)$ is the 2-D proton beam distribution. The response matrix can be estimated based on the MC simulation.

\begin{figure}[htbp]
\begin{center}
\begin{overpic}[width=7.5cm,height=5.5cm,angle=0]{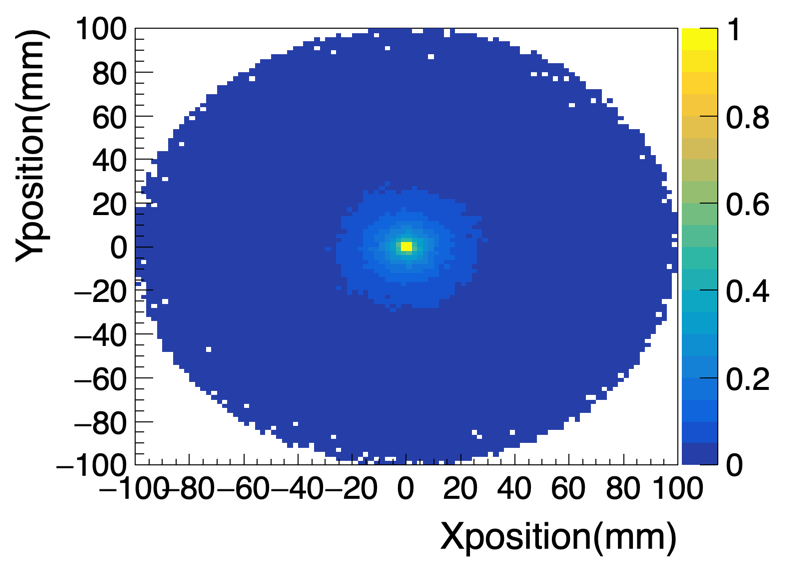}
\end{overpic}
\begin{overpic}[width=7.5cm,height=5.5cm,angle=0]{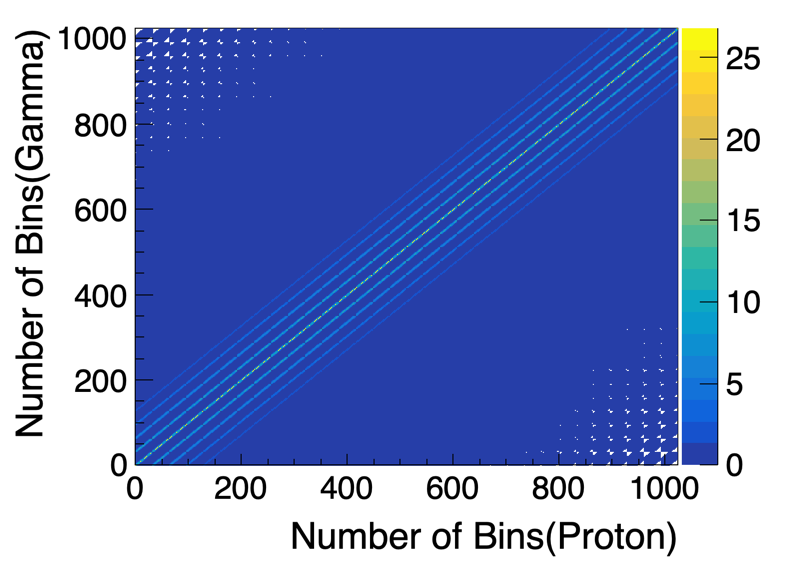}
\end{overpic}
\end{center}
\caption{(Left) Distribution of gamma from target surface generated by incident proton beam of point source. (Right) The 2-D distribution of the response matrix.}
\label{FitToGammaDistribution}
\end{figure}

\begin{figure}[htbp]
\begin{center}
\begin{overpic}[width=6.5cm,height=5.0cm,angle=0]{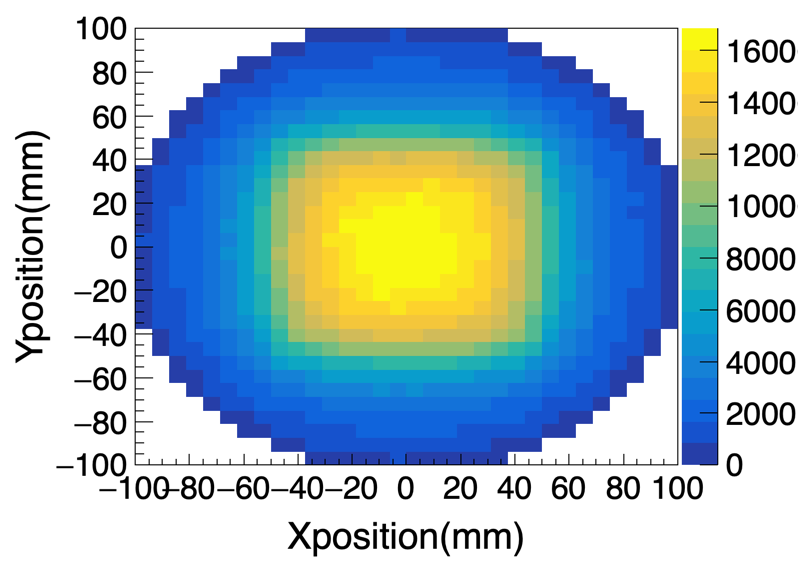}
\end{overpic}
\begin{overpic}[width=6.5cm,height=5.0cm,angle=0]{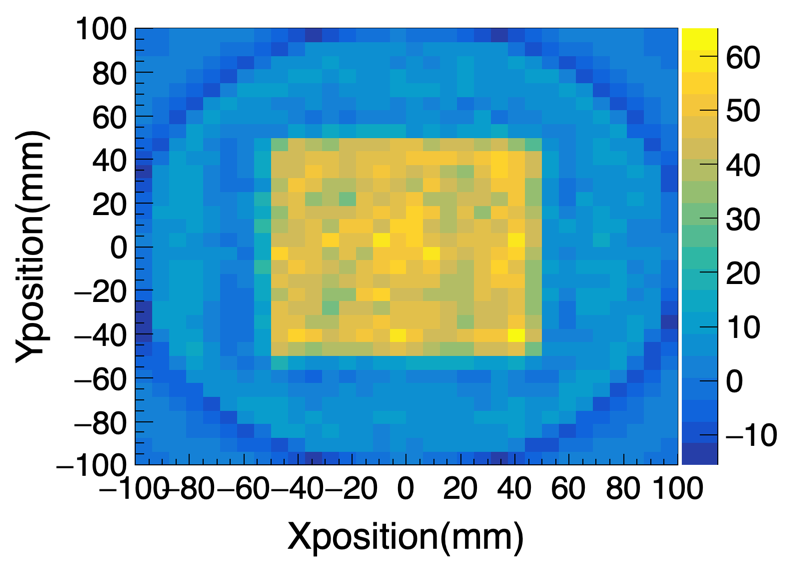}
\end{overpic}
\end{center}
\caption{(Left) Distribution of gamma with 32$\times$32 pixels from the target surface. (Right) The reconstructed distribution of proton with 32$\times$32 pixels obtained by solving linear equation based on the iterative method.}
\label{ReconstructedProtonImage}
\end{figure}

The physical process is simulated with GEANT4~\cite{Geant4} by impinging 1.6 GeV point-like proton beam onto a tungsten target. Figure~\ref{FitToGammaDistribution} (Left) is the gamma distribution detected on the target surface from the point-like proton beam. In order to calculate the response matrix, the gamma distribution is described by 2-D double Gaussian function. The 2-D distribution of the response matrix is drawn in Fig.~\ref{FitToGammaDistribution} (Right). Figure~\ref{ReconstructedProtonImage} (Left) shows the gamma distribution of the target surface from the rectangular proton beam (10 cm$\times$10 cm). The gamma distribution consists of 1024 pixels (32$\times$32). The linear equation could be solved by algebraic reconstruction technique (ART) to reconstruct the proton beam distribution in Fig.~\ref{ReconstructedProtonImage} (Right)~\cite{ARTMethod}. The reconstructed proton distribution has a clear circle, which is the boundary of the cylindrical target.


\begin{figure}[htbp]
\begin{center}
\begin{overpic}[width=12.0cm,height=8.cm,angle=0]{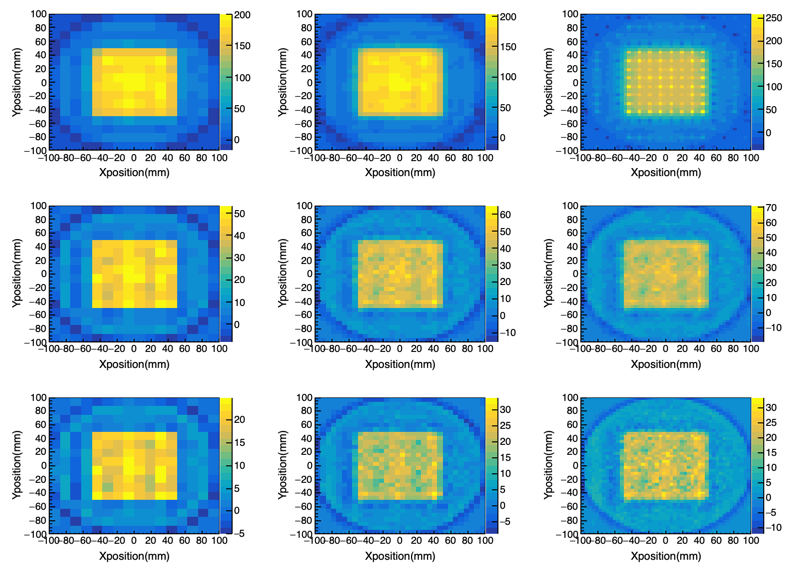}
\end{overpic}
\end{center}
\caption{Reconstructing distribution of proton beam spot based on measuring gamma distributions with different combinations of the pixelated CZT. The column is the number of bins for reconstructing distribution of proton beam spot (First column [16, 16], second column [32, 32], third column [48, 48]). The first row proton distributions results from measuring gamma distribution with 16$\times$16 pixels, the second row from 32$\times$32 pixels, the third row from 48$\times$48 pixels.}
\label{ReconstructionImageOfCZT}
\end{figure}

The size of the CZT is 25.4 mm$\times$25.4 mm$\times$5 mm, and the number of anode readout pixels is 256 (16$\times$16). Taking into account the different assembly combinations of the pixelated CZT, 4 pieces of the same CZT consist of 32$\times$32 pixels and 9 pieces have 48$\times$48 pixels. In order to obtaining the best reconstructed proton distribution, we have performed different bins of gamma distribution and proton distribution. The reconstructed proton distributions are shown in Fig.~\ref{ReconstructionImageOfCZT}. Its columns are the number of bins for reconstructing image of proton beam spot. The first-row distributions of reconstructing proton result from the measuring gamma distribution with 16$\times$16 pixels. The second-row distributions result from the measuring gamma distribution with 32$\times$32 pixels. The third-row distributions result from the measuring gamma distribution with 48$\times$48 pixels. With the increase of pixels, the reconstruction of proton distribution has a much higher resolution. Although a single detector was used for pinhole imaging, the distribution of proton on the target surface has been also reconstructed clearly.

\subsection{Neutron imaging}

Because we need distinguish neutrons from the secondary gammas before performing neutron imaging, it is difficult to measure the distribution of neutron from pinhole imaging directly. $^{10}$B has a high cross section for thermal neutron absorption. The probability of emitting 478 keV $\gamma$-rays is about 94$\%$ in the $^{10}$B(n, $\alpha$) reactions. The neutron distribution could be reconstructed with Compton imaging method by measuring the characteristic gamma distribution from the $^{10}$B(n, $\alpha$) reactions. The CZT detector is sensitive to the emitting characteristic gamma of 478 keV. The pixelated CZT detector combined with the neutron converter can be used for neutron imaging. Making use of the advantage of the characteristic (478 keV) gamma could suppress the backgrounds of gamma and improve the quality of Compton imaging.

\begin{figure}[htbp]
\begin{center}
\begin{overpic}[width=7.5cm,height=5.5cm,angle=0]{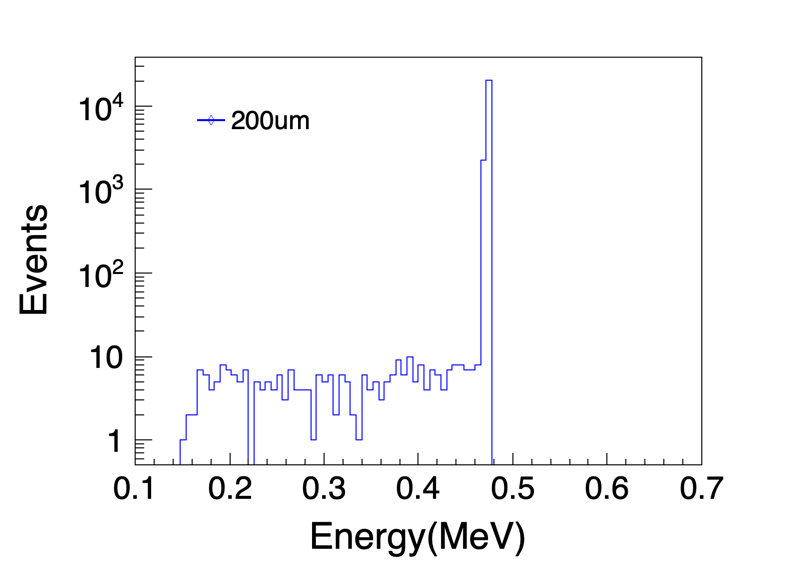}
\end{overpic}
\begin{overpic}[width=7.5cm,height=5.5cm,angle=0]{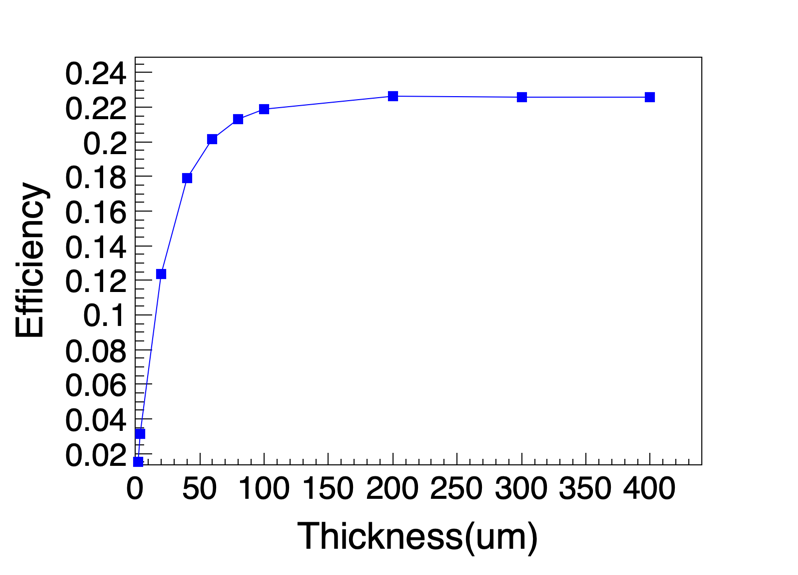}
\end{overpic}
\end{center}
\caption{(Left) Energy distribution of gamma from $^{10}$B(n, $\alpha$) reactions. (Right) Optimization of the detected efficiency versus the thickness of $^{10}$B based on the characteristic gamma.}
\label{OptimizationOfThickness}
\end{figure}

Taking the detection efficiency into account, we performed the optimization to the thickness of $^{10}$B corresponding to the thermal neutron. Figure~\ref{OptimizationOfThickness} (Left) shows the energy distribution of secondary gammas from $^{10}$B(n, $\alpha$) reactions. The efficiency curve of detecting gamma is drawn in Fig.~\ref{OptimizationOfThickness} (Right). The efficiency is calculated by requiring the energy range of gamma (larger than 0.47 MeV). The uncertainty of efficiency is determined according to the following formula

\begin{center}
\begin{equation}
\sigma_{\epsilon}=  \sqrt{\frac{\epsilon(1-\epsilon)}{n}},
\label{ErrOfEpsilon}
\end{equation}
\end{center}

\begin{figure}[htbp]
\begin{center}
\begin{overpic}[width=4.5cm,height=4.0cm,angle=0]{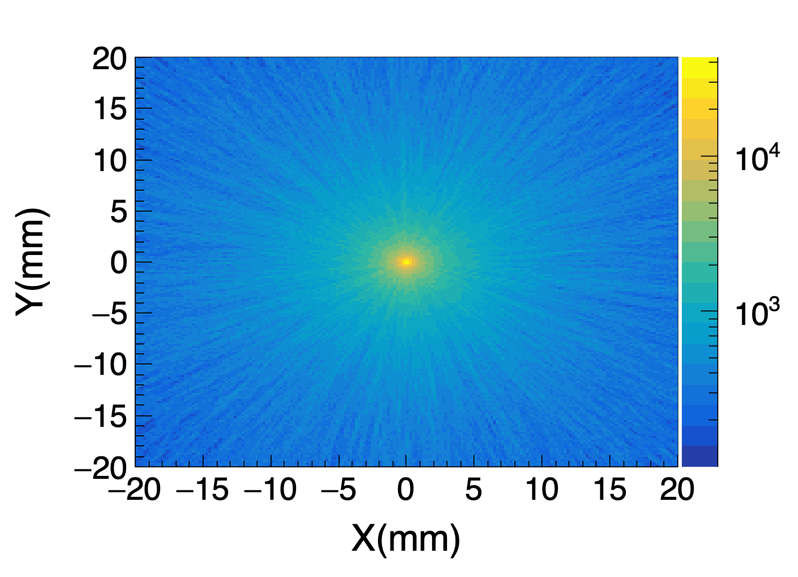}
\end{overpic}
\begin{overpic}[width=4.5cm,height=4.0cm,angle=0]{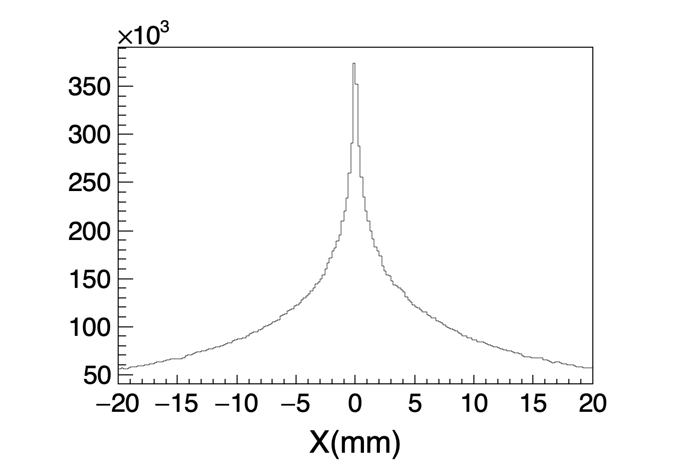}
\end{overpic}
\begin{overpic}[width=4.5cm,height=4.0cm,angle=0]{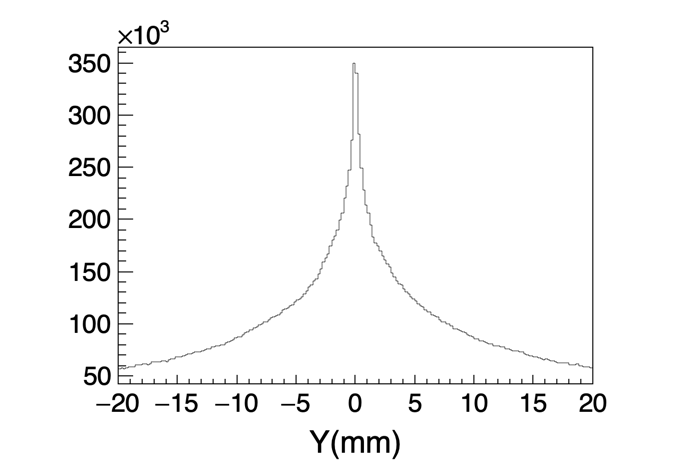}
\end{overpic}
\end{center}
\caption{(Left) Result of Compton imaging from the point-like gamma source. (Middle) X-axis projection of Compton imaging. (Right) Y-axis projection of Compton imaging.}
\label{PointDistributionNeutronImaging}
\end{figure}

The neutron imaging is also reconstructed by using a similar method of solving the linear equation~\ref{LinearEquation}. The response matrix was also calculated based on the gamma distribution of Compton imaging from the point-like gamma source. The 2-D gamma distribution and 1-D projections of Compton imaging are shown in Fig.~\ref{PointDistributionNeutronImaging}. We used the physical list of QGSP-BIC-HP to simulate a square source with the side length of 10 mm emitting the isotropic $\gamma$-rays based on GEANT4. The reconstruction of initial gamma distribution was implemented based on Compton imaging. Figure ~\ref{PointDistributionNeutronImaging} (Left) and (Right)  shows the comparison between the raw and the reconstructed square distribution of characteristic gamma. The secondary neutron distribution from pinhole imaging corresponds to the vertex distribution of characteristic (478 keV) gamma.

\begin{figure}[htbp]
\begin{center}
\begin{overpic}[width=4.5cm,height=4.0cm,angle=0]{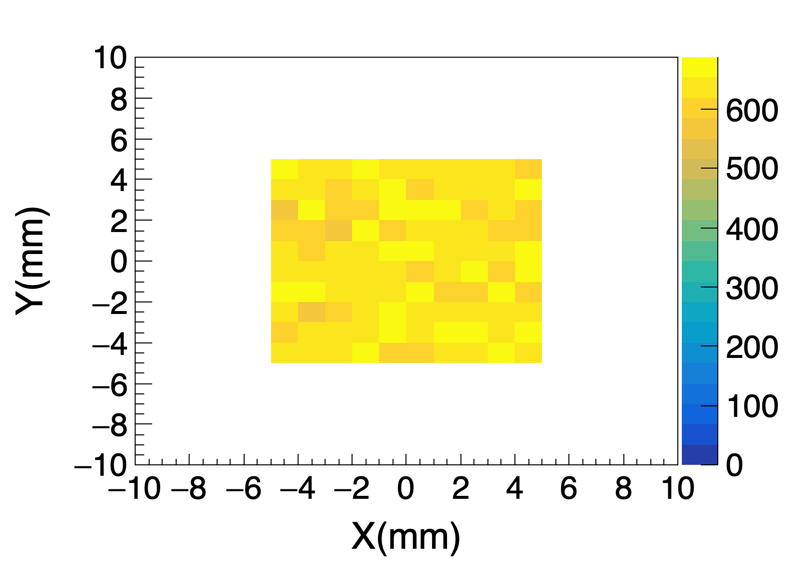}
\end{overpic}
\begin{overpic}[width=4.5cm,height=4.0cm,angle=0]{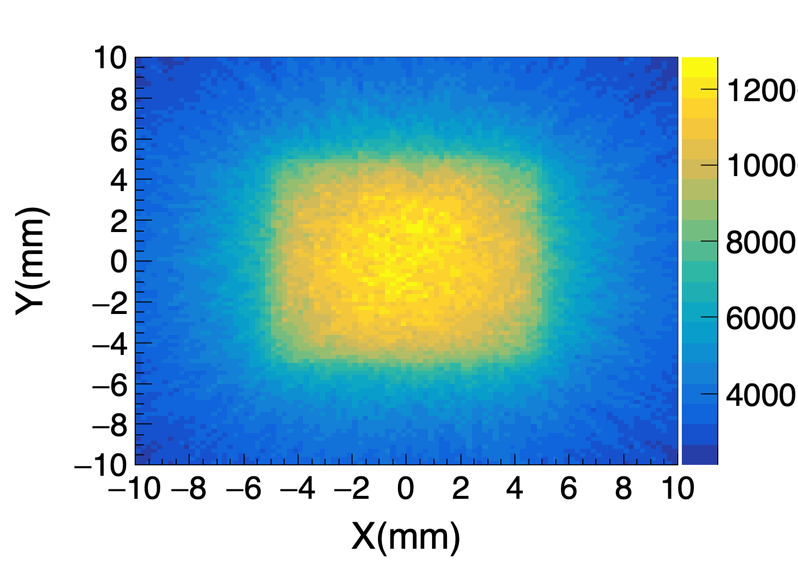}
\end{overpic}
\begin{overpic}[width=4.5cm,height=4.0cm,angle=0]{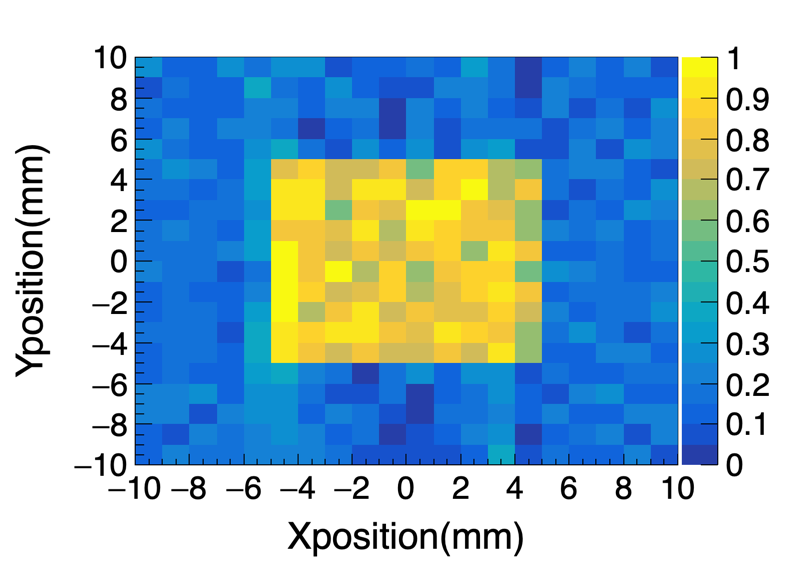}
\end{overpic}
\end{center}
\caption{(Left) Original square distribution of the characteristic (478 keV) gamma. (Middle) The result of Compton imaging for the square distribution of the characteristic (478 keV) gamma. (Right) The reconstructed gamma distribution of the characteristic (478 keV) gamma.}
\label{RecNeutronImaging}
\end{figure}


\section{Conclusion}

The pixelated detector is suitable for measuring the gamma distribution based on the principle of pinhole imaging and Compton imaging. Pinhole imaging has a good application prospect for the measurement of high intensity proton beam spot. Compton imaging method used for the measurement of the neutron distribution is proposed for the first time in this work. A 3-D position-sensitive CZT detector can work at room temperature, which has high detection efficiency and energy resolution for gamma. Based on the pixelated CZT gamma detector, the system of Compton imaging will be applied to measure the proton beam spot at CSNS in the future.

\section{Acknowledgement}

This work was supported by the National Natural Science Foundation of China (Project: 12075135) and the China Postdoctoral Science Foundation (No. 2021M691859).



\vspace{-1mm}
\centerline{\rule{80mm}{0.1pt}}
\vspace{2mm}

\end{document}